\documentclass[11pt,a4paper]{article}
\usepackage[makeroom]{cancel}
\usepackage{bm}
\usepackage{amsmath}
\usepackage{amsfonts}
\usepackage{amssymb}
\usepackage{graphicx}
\usepackage{nicefrac}
\usepackage{authblk}
\usepackage{cite}
\usepackage[left=2.54cm,top=2.54cm,right=2.54cm,bottom=2.54cm,nohead]{geometry}
\DeclareGraphicsExtensions{.png,.jpg,.pdf}
\usepackage{epstopdf}
\usepackage{float}
\usepackage{fouriernc} 
\usepackage{enumitem}
\usepackage[utf8]{inputenc}
\usepackage[english]{babel}
\usepackage{mathtools}
\usepackage{amscd}
\usepackage{mathrsfs}
\usepackage{amsthm}
\usepackage{siunitx}
\usepackage{caption}
\usepackage{subcaption}
\setlist{nolistsep,leftmargin=*}

\usepackage{mhchem} 
\usepackage{cleveref} 
\usepackage{xcolor}

\title{Electrostatic cooling at electrolyte-electrolyte junctions}

\makeatletter

\renewcommand\AB@authnote[1]{\textsuperscript{\normalfont#1}}

\makeatother

\author[1,2]{S. Porada,}
\author[1]{H.V.M. Hamelers,}
\author[1]{P.M. Biesheuvel}

\affil[1]{Wetsus, European Centre of Excellence for Sustainable Water Technology, Leeuwarden, The Netherlands.}
\affil[2]{Membrane Science and Technology, University of Twente, The Netherlands.}

\date{} 

\newcommand{\temphl}[1]{{\color{black} #1}} 

\begin{document}

\maketitle

\begin{abstract}

Electrostatic cooling is known to occur in conductors 
and in porous electrodes in contact with aqueous electrolytes. 
Here we present for the first time evidence of electrostatic cooling 
at the junction of two electrolyte phases. These 
are, first, water containing salt, and, second, an ion-exchange membrane, which is a water-filled porous layer containing a large concentration of fixed charges. When ionic current is directed through such a membrane in contact with aqueous phases on both sides, a temperature difference develops across the membrane which rapidly switches sign when the current direction is reversed. The temperature difference develops because one water-membrane junction cools down, while the other heats up. Cooling takes place when the inner product of ionic current $\textbf{I}$ and field strength $\textbf{E}$ is a negative quantity, which is possible in 
the electrical double layers that form on the surface of the membrane. Theory reproduces the magnitude of the effect but overestimates the rate by which the temperature difference across the membrane adjusts itself to a reversal in current. 

\end{abstract} 


\begin{figure}[H]
\centering
\includegraphics[width=0.35\textwidth]{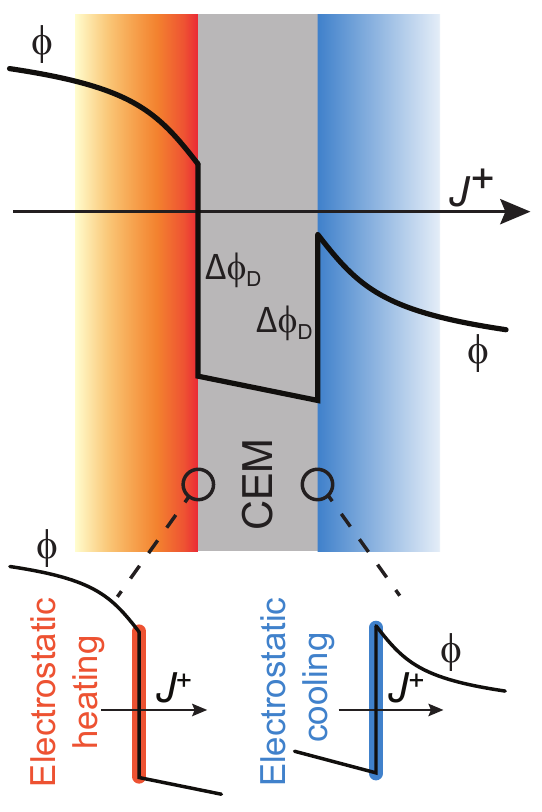}
\caption{A temperature difference develops across a cation-exchange membrane (CEM) 
when ionic current runs 
through the membrane. At almost all locations the electrical potential, $\phi$, decreases in the direction of current, leading to Joule heating because of an ionic resistance, but there is cooling in one of the Donnan regions (electrical double layers) 
located on the outsides of the membrane.} 
\label{fig1}
\end{figure}

\newpage

``There are few facts in science more interesting than those which establish a connexion between heat and electricity,'' wrote James Prescott Joule in 1841~\cite{Joule_1841}, and this is possibly as true today as it was the case then. 

Joule heating is the 
irreversible electrostatic heating described by $\textbf{I}^2 R$ where $\textbf{I}$ is the current density (in A/m$^2$) and $R$ a local resistance (in $\Omega$m)~\cite{Domenicali_1954,Chu_1959}. Joule heating is always positive: only heating is predicted to occur. However, cooling due to current is possible as well, for instance at the junction of two materials that allow current to pass. Examples are the contact point of two thermoelectric materials (the Peltier effect)~\cite{DiSalvo_1999,Bell_2008}, and the electrical double layer (EDL) in porous electrodes~\cite{Schiffer_2006,Dandeville_2011,Janssen_2017_118,Janssen_2017,Atlas_2018,Munteshari_2018}. 
Electrostatic heating and cooling in EDLs are both described by the inner product of $\textbf{I}$ and field strength $\textbf{E}$, a term which can be both positive and negative~\cite{Janssen_2017,Atlas_2018,Munteshari_2018,DeGroot_1951,Landau_1960
,Katchalsky_1965,Hartig_1993,Kharkats_1998,Gu_2000,Robson_2003,
Kontturi_2008,Biesheuvel_2014,Janssen_thesis}. When Ohm's law applies and thus current flows in the direction of the field strength, then the classical expression $\textbf{I}^2R$ results (Joule heating), but when additional driving forces act on the charge carriers, the electric power density $\textbf{I}\cdot \textbf{E}$ does not directly depend on the resistance $R$ and can also be negative, which implies local cooling. 
These additional forces operate for instance 
in 
porous electrodes, 
plasmas~\cite{Hartig_1993,Robson_2003}, semiconductor devices~\cite{Wachutka_1990}, the Earth's 
ionosphere~\cite{Kosch_1995,Zhang_2005,Zhu_2005} and magnetosphere~\cite{Graham_2014}, and in other examples of 
thermoelectric effects~\cite{Domenicali_1954,Chu_1959}. 
Though the formulation of electric power density as the inner product of \textbf{I} and \textbf{E} can be encountered in the abovementioned sources, there is not much, if any, corroboration that it indeed can lead to electrostatic cooling, except for experiments involving (semi-)metallic conductors and electrodes.

In this Letter we show for the first time evidence of electrostatic cooling in an electrolyte, i.e., a phase with ions as charge carriers, not in contact with any electron-conducting interface. 
In our experiments we measure the temperature development on both sides of an ion-exchange membrane (IEM) in contact with a salt solution, see Fig.~\ref{fig1}. In this system, we have two electrolyte-electrolyte junctions very near one another, one on each side of the IEM. An IEM is a polymeric gel-like layer that is permeable to ions and water and contains a very high concentration of fixed charges, bound to the polymer network (more than $X=5$ M of fixed charges per volume of water for commercial IEMs). When the membrane contains fixed negative charge, it is highly permeable to cations, thus is a cation-exchange (or cation-selective) membrane (CEM). When ionic current runs through such an IEM, in most regions the electric power density is positive, namely wherever the current runs towards lower electrical potentials, and at all these positions heat is produced. However, in one of the EDLs, on one face of the membrane, the current runs to a higher electrical potential, and thus the electric power density is negative. The local cooling in this EDL, while at the same time the other EDL heats up, should lead to a temperature difference across the membrane, $\Delta T$, 
which changes sign when the current direction is reversed. 
We aim to measure the temperature developments on each side of a CEM with current flow, and present theory to describe these temperature changes. 


That electrostatic cooling occurs somewhere in an electrolyte-membrane system, can 
be inferred on conceptual grounds from considering the technology called ``osmotic power'' or ``blue energy''~\cite{Rica_2012,Siria_2013,Tedesco_2017}, a method in which electrical power is extracted from the controlled mixing of water of high and low salinity, for instance by the use of many parallel IEMs in reverse electrodialysis (RED). Assuming two ideal salt solutions, with different salinity but the same temperature, when they are mixed without extraction of power, the temperature must stay the same (assuming no heating due to fluid friction). When, instead, electrical energy is extracted in an RED device that is fed with these two solutions, then the solutions that leave the device must have become colder than the inlet streams~\cite{Janssen_thesis}. This is because when the extracted electrical energy is used to later heat up these exit streams, they must (on average) end up at the same temperature as in the case of no extraction of energy (namely, the same as the temperature of the inlet streams). Therefore, prior to being heated by the generated electrical energy, the solutions that leave the RED device must have become colder than the inlet streams.  
The question then is, where does this cooling take place? 
In ref.~\cite{Biesheuvel_2014} it was suggested that this happens on one of the surfaces of the IEMs, always on the side of the low-salinity solution, because at this surface the term $\textbf{I} \cdot \textbf{E}$ is negative in the EDL. In  this Letter we provide indirect support for this prediction in our combined experimental and theoretical study of such temperature effects in an IEM in contact with aqueous solutions. Our experiment, however, is not based on a spontaneous process such as RED but was done with solutions of 0.5 M KCl in water on both sides of the membrane. The advantage is that with equal salt concentrations on each side, the experiment becomes more symmetric, with the temperatures on each side of the membrane expected to respond in the same way after each current reversal, irrespective of the direction of current. 

\begin{figure}[H]
\centering
\includegraphics[width=0.4\textwidth]{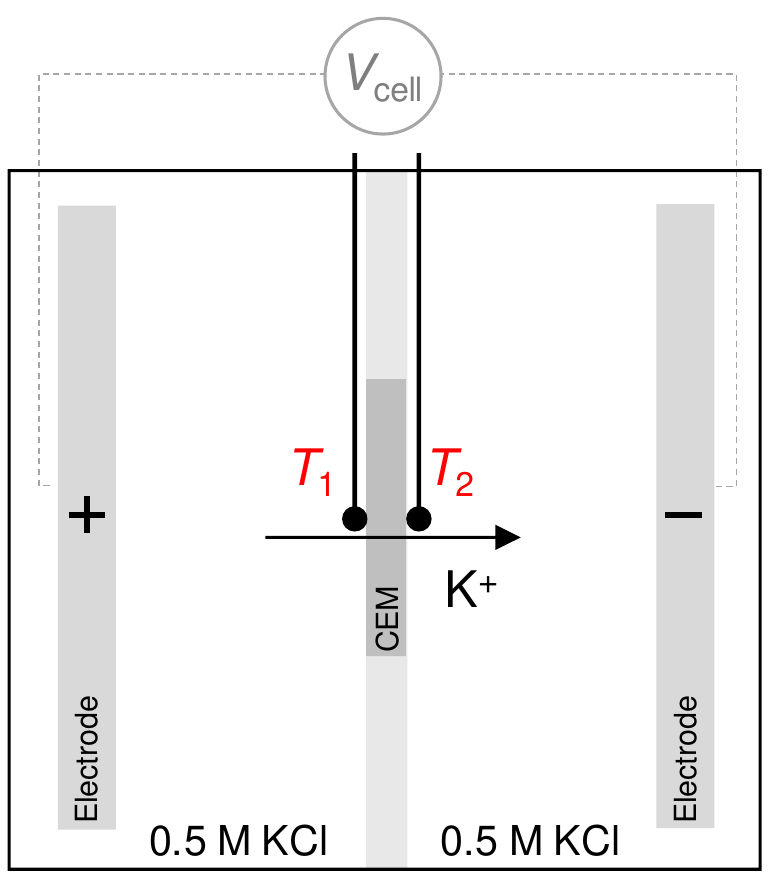}
\caption{\temphl{Schematic of the experimental setup to measure temperature differences across an ion-exchange membrane in response to an applied current.}} \label{fig_setup}
\end{figure}

Experiments were performed at $T_\text{room}$ with a CEM prepared from a cationic ionomer solution (FKS solution, 16 wt\% of polymer dissolved in NMP; Fumatech, Germany; thickness $\sim 200$ $\mu$m; density of fixed negative charge $X=4.5\pm 0.5$ M). The membrane is placed in a {\O} 0.95 cm circular hole in between two cylindrical compartments ({\O} 2 cm, length $\sim 6$ cm) with electrodes on either side, \temphl{see Fig.~\ref{fig_setup}}. 
Near both electrodes the compartments are open to air to allow gases that develop at the electrodes to escape. During an experiment there is no flow of water. Temperature detectors (miniature RTD Sensor-Pt100, TC Direct, UK; connected to a PT-104 platinum resistance data logger, Pico Technology, UK) were placed on each side of the membrane near the center of the circular membrane area, contacting the membrane. The wires are 0.5 mm in 
diameter, with the temperature detector placed on the very tip. 
Experiments are done at various currents, with the switching of current every 60 or 90 s. 
Reported values for current density refer to the area of the membrane. 
\temphl{Experiments with uncharged membranes are discussed at the end of this report.}

\begin{figure}[H]
\centering
\includegraphics[width=0.8\textwidth]{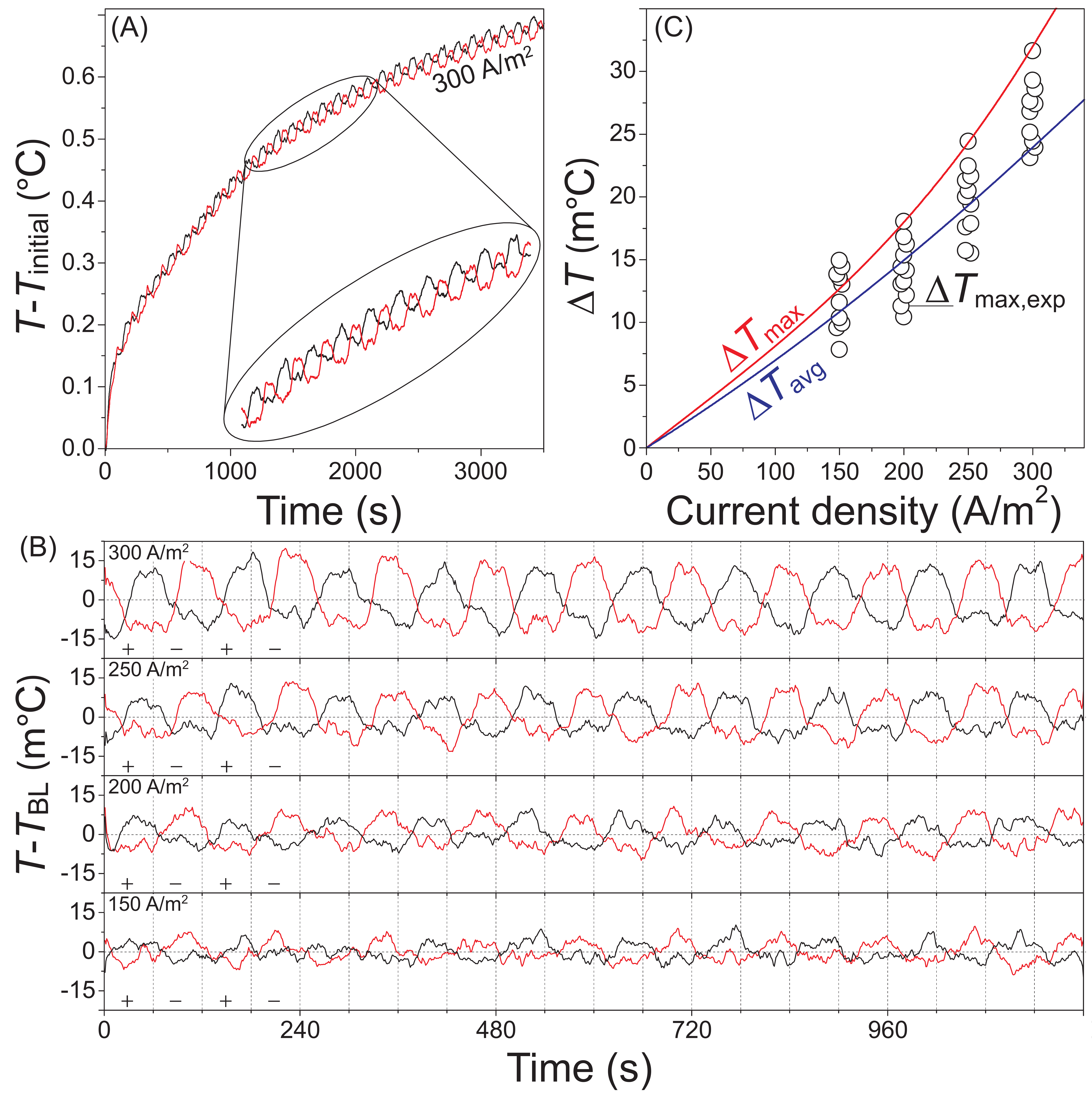}
\caption{The temperatures, $T$, measured on each side of an ion-exchange membrane
placed between two 0.5 M KCl solutions (A) gradually go up with time $t$, (B) but on a shorter time scale, with every 60 s the current direction being reversed, the two $T$-$t$ traces reverse, with one membrane face now starting to cool down \temphl{, the other heating up}. (C) The temperature difference between the two sides, $\Delta T$, increases with current.} \label{fig2}
\end{figure}

\begin{figure}[H]
\centering
\includegraphics[width=0.85\textwidth]{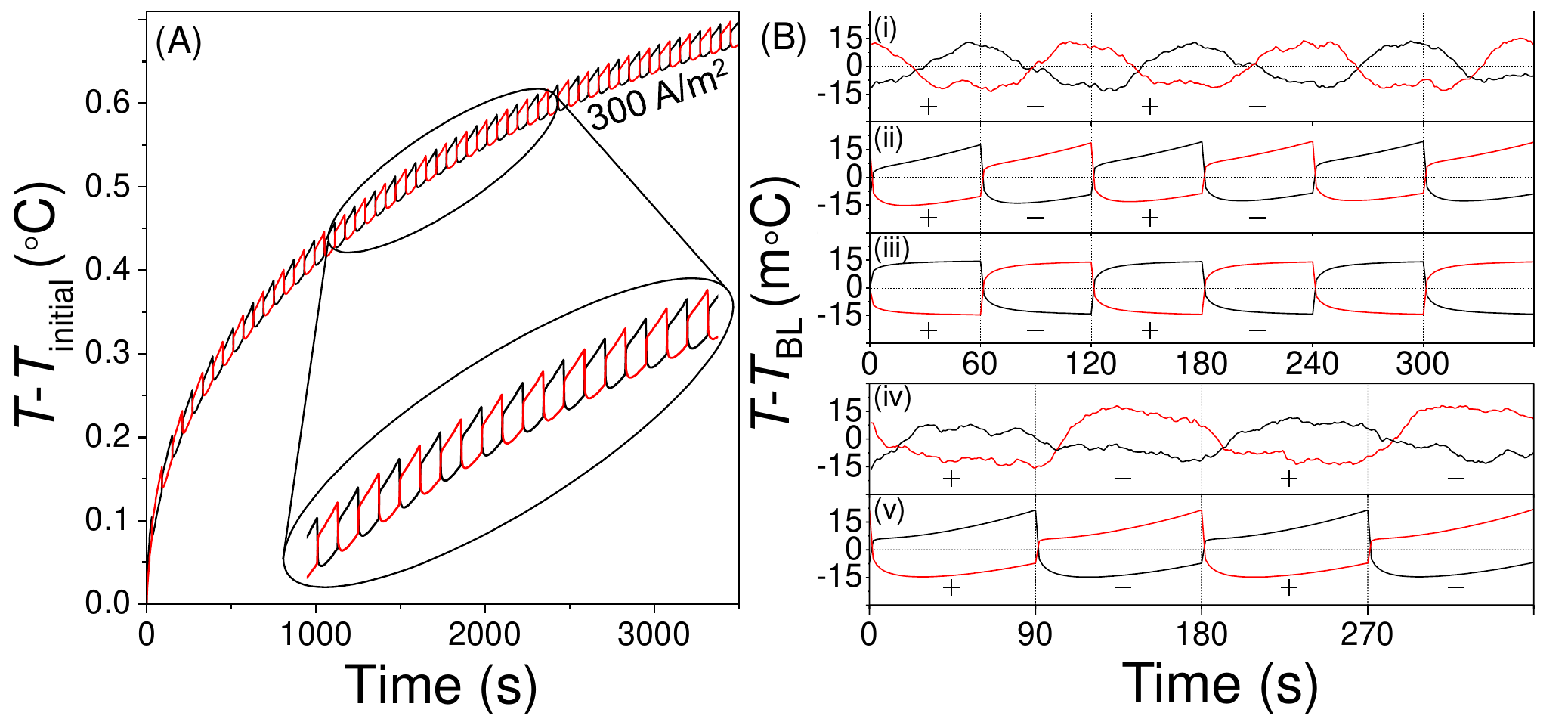}
\caption{(A) Theoretical calculations for the temperature increase of each membrane interface over time. (B) Data (i and iv) and theory (ii, iii, v) for the temperature traces after subtracting the baseline temperature, for 60 s and 90 s switching times.} \label{fig3}
\end{figure}

As Fig.~\ref{fig2}A shows, upon applying a current (switched every 60 s), the two measured temperatures (one on each side of the membrane) steadily increase over time due to Joule heating by the ionic resistances in water and membrane. Because of heat loss, for instance along the sides of the 6 cm long compartments, the temperatures gradually level off.  Zooming in on individual 60 s-periods, we see that upon current reversal one measured temperature decreases, while the other increases. 
We average the two temperature-time traces and fit a second order polynomial to obtain the baseline temperature, $T_\text{BL}$. After subtracting $T_\text{BL}$ from the temperature signal, we obtain the temperature traces in Fig.~\ref{fig2}B which show how during each 60 s-period one temperature decreases, and the other increases, a trend which is reversed upon changing the current. For each current density, this experiment is done for about 1 hr (60 reversals) and we consistently see that the measured temperatures display the counter-cyclic behavior as shown in Fig.~\ref{fig2}B. We can measure the maximum temperature difference in each period, $\Delta T_\text{max}$, and plot these versus current in Fig.~\ref{fig2}C which shows how $\Delta T_\text{max}$ is roughly proportional with current. As Fig.~\ref{fig2}B suggests, after 60 s the temperature changes within the 60 s-period have stabilized, an observation supported by similar data with switching times of 90 s, which more clearly show that after 30 to 60 s the temperature signal levels off (fig~\ref{fig3}B, 4$^\text{th}$ row).

We construct a theoretical model based on ion mass balances and a thermal energy balance (also called heat equation). For the salt solutions (on each side of the membrane), ion transport is described by the Nernst-Planck equation, and assuming electroneutrality 
and equal ion diffusion coefficients (well approximated for KCl, with $D \sim 2.0\cdot 10^{-9}$ m$^2$/s), we arrive at the salt mass balance~\cite{Tedesco_2017}
\begin{equation}
{\partial c}/{\partial t}=D \nabla^2 c
\label{eq1}
\end{equation}
where $c$ is salt concentration, and $t$ is time. The heat equation is given by~\cite{DeGroot_1951,Kharkats_1998,Biesheuvel_2014,Janssen_2017_118,Janssen_2017}
\begin{equation}
\rho c_\text{p}\hspace{1mm}{\partial T}/{\partial t}=\nabla \cdot \left( \lambda \nabla T \right) 
+\textbf{I} \cdot \textbf{E}
\label{eq2}
\end{equation}
where $\rho c_\text{p}$ is the heat capacity (of water, $\rho c_\text{p}=4.2$ MJ/m$^3$/K) 
and $\lambda$ the thermal conductivity (for water $\lambda=0.6$ W/m/K). The field strength is $\textbf{E}=-V_\text{T}\nabla \phi$ with $V_\text{T}=k_\text{B}T/e$ and $\phi$ is the dimensionless electrical potential. For the assumptions underlying Eq.~\eqref{eq1}, current density \textbf{I} relates to $\phi$ according to $\textbf{I}=-2DcF\nabla\phi$~\cite{Tedesco_2017}. 
Eq.~\eqref{eq2} is also used in the membrane but with 
$\rho c_\text{p}=2.1$ MJ/m$^3$/K (based on 30 vol\% water in the membrane and $\rho c_\text{p}$ of the polymer 30\% of that of water). We assume a membrane that is perfectly selective to only allow cations to go through, and no anions. Thus in the membrane we can assume an ion concentration equal to $X$, and thus in the membrane we only have migration (Ohm's law) and we can replace $\textbf{I}\cdot \textbf{E}$ in Eq.~\eqref{eq2} by $I^2\hspace{1mm}R_\text{m}$ with $R_\text{m}=V_\text{T}/\left(XFD\right)$. At each membrane-water edge, we have Donnan equilibrium, where the Donnan potential, $\Delta \phi_\text{D}$ (potential in membrane minus outside), is given by 
\begin{equation}
\Delta \phi_\text{D} = - \sinh^{-1} \left({X}/{2 c^*}\right)
\label{eq3}
\end{equation}
where $c^*$ is the salt concentration in the water next to the membrane, just outside the EDL, as obtained from Eq.~\eqref{eq1}. Eq.~\eqref{eq3} assumes an ideal salt solution with ion statistics in the EDL described by Boltzmann's law. \temphl{For a very thin layer, such as an EDL, where gradients in potential are strong, and we expect strong changes in the temperature gradient, we can evaluate Eq.~\eqref{eq2} by neglecting the accumulation term on the left hand side.} We can then integrate Eq.~\eqref{eq2} across the membrane-water interface, i.e., across the EDL, which we can to be planar, to result in
\begin{equation}
\left(\vphantom{\nabla^2_2}\lambda \hspace{1mm}\nabla T \cdot \textbf{n} \right)_\text{in}-\left(\vphantom{\nabla^2_2}\lambda \hspace{1mm}\nabla T \cdot \textbf{n} \right)_\text{out} = \textbf{I} \cdot \textbf{n} \hspace{1mm} V_\text{T} \Delta \phi_\text{D}
\label{eq4}
\end{equation}
with $\textbf{n}$ the unit vector directed at right angles into the membrane. \temphl{Eq.~\eqref{eq4} is a new result and describes both electrostatic heating and cooling in a(n infinitely) thin region with non-negligible voltage drop, $\Delta\phi_\text{D}$. This reversible heating or cooling is solely due to running a current through a Donnan voltage step as found in EDLs at charged interfaces, and is thus totally separate from Joule heating, which relates to a resistance to current across a layer of non-vanishing thickness.} \temphl{Note that across the EDL (i.e., from one end to the other) the temperature is continuous, and only its gradient makes a step change between two points on either side of the EDL, as described by Eq.~\eqref{eq4}. This is different for potential, $\phi$, and ion concentrations, $c_i$, which both have different values on either side of the EDL.}

To empirically include heat loss along the sides of the two cylindrical compartments that are on each side of the membrane, we include on the right-hand side of Eq.~\eqref{eq2} a term $- \alpha \left(T-T_\text{ext}\right)$, where $\alpha$ is a heat exchange coefficient 
and $T_\text{ext}$ is an external temperature 
(we do not include this term for the membrane).

We solve this set of equations dynamically, in a one-dimensional geometry. For the membrane, Eq.~\eqref{eq2} is solved in a Cartesian geometry (coordinate axis directed straight across the membrane as in Fig.~\ref{fig1}). For the solution compartments, we likewise consider two straight sections, but much longer, adjacent to each electrode ($L=5.4$ cm, {\O} 2 cm), while we include in the calculation two short regions with a spherical geometry right next to the membrane, i.e., located between the long sections and the membrane ($r_\text{inner}=6.21$ mm, $r_\text{outer}=13.07$ mm). This allows for a smooth transition from the smaller membrane ({\O} 0.95 cm) to the wider compartment ({\O} 2 cm) while retaining a 1D model formulation. Eq.~\eqref{eq4} is solved at the two interfaces between the membrane and the spherical sections, with $\textbf{I}\cdot \textbf{n}$ replaced by the current density \textit{I} that runs through the membrane. Note that \textit{I} decreases through the spherical section and is $>4\times$ lower in the outer compartments than in the membrane. The external temperature, $T_\text{ext}$, is set equal to the initial temperature in the system.

Theoretical calculations for the gradual temperature change over many cycles can be fit to the data in Fig~\ref{fig2}A by using appropriate values for $\alpha$ and $R_\text{m}$ (the average of the two theoretical $T-t$ traces in Fig.~\ref{fig3}A exactly matches the experimental average in Fig.~\ref{fig2}A). We use $\alpha=1.5$ kW/m$^3$/K and $R_\text{m}=1.30$ $\Omega$m, which implies a cation diffusion coefficient in the membrane $\sim 45 \times$ reduced compared to the value in solution (similar to a reduction factor derived in ref.~\cite{Tedesco_2017}). This membrane resistance is $\sim 9$ times larger than in the salt solution of $c=0.5$ M. The temperature difference that develops between the two membrane surfaces can be reproduced by using an optimized value for the thermal conductivity in the membrane of $\lambda_\text{m}=0.13$ W/m/K (similar to values in ref.~\cite{Bock_2019}), see Fig.~\ref{fig2}C, where 
experimental data for $\Delta T_\text{max}$ in a 60 s-period are compared with the theoretically obtained values for $\Delta T_\text{max}$, and for $\Delta T_\text{avg}$, the average temperature difference in the 60 s-period. 

However, despite this good fit, an important discrepancy between theory and data is quite apparent, which is the rate of temperature change after the current  
is reversed. This is much too fast in the theory, see Fig.~\ref{fig3}B, where the traces for data and theory do not compare too well with respect to the rate of temperature change. To improve this situation, we tested various model adjustments but none of them improved the fit. We first made a simplified calculation in which the salt concentration in the solutions is assumed to stay at $c=0.5$ M. This model fits data equally well, see Fig.~\ref{fig3}B, 3$^\text{rd}$ trace (fit with 
$R_\text{m}=1.83$ $\Omega$m and $\lambda_\text{m}=0.10$ W/mK), but again the temperature changes too quickly after the current is reversed. The assumption of an unvarying value of the salt concentration is correct for a non-selective membrane where the anion and cations have equally large fluxes but flow in opposite directions, which theoretically requires an uncharged membrane and anion and cation diffusion coefficients that are equal to one another, both in solution and in the membrane. In reality, membrane selectivity will be in between the limit of perfect selectivity (on which all our other calculations are based) and the limit of zero selectivity.

\newpage

Other modifications of the model also did not improve the situation. 
First, we considered that the temperature detectors have a diameter of 0.5 mm and thus the temperature is likely measured not at the membrane surface directly, but at $\sim0.25$ mm away from the membrane. Indeed the temperature changes more slowly at this distance (as well as further away), but the amplitude of the changes in temperature also dramatically decreases, and thus $\Delta T$ also goes down. This cannot be remedied in the theory by adjustment of any of the fit parameters. Second, we considered that there is advection of heat through the membrane. This may play a role because water flows with the cation current through the membrane. At a current density of $I=300$ A/m$^2$ this volume flow is estimated for this membrane at $\sim 0.3-0.4$ $\mu$m/s~\cite{Porada_2018}. Assigning a $\rho c_\text{p}$ of water to that flow and including advection in Eq.~\eqref{eq2} (within a model with fixed solution salt concentration, like for the 3$^\text{rd}$ trace in Fig.~\ref{fig3}B), however, hardly changed the predicted rate of temperature change. Setting up a full two-dimensional model geometry is certainly more accurate than our present 1D model, but it is not obvious why such a model refinement would make the calculated temperature on the membrane faces change more gradually. Thus it remains unknown for now why there is this discrepancy between theory and data with respect to the rate of temperature change after the current is reversed. 

\begin{figure}[H]
\centering
\includegraphics[width=0.85\textwidth]{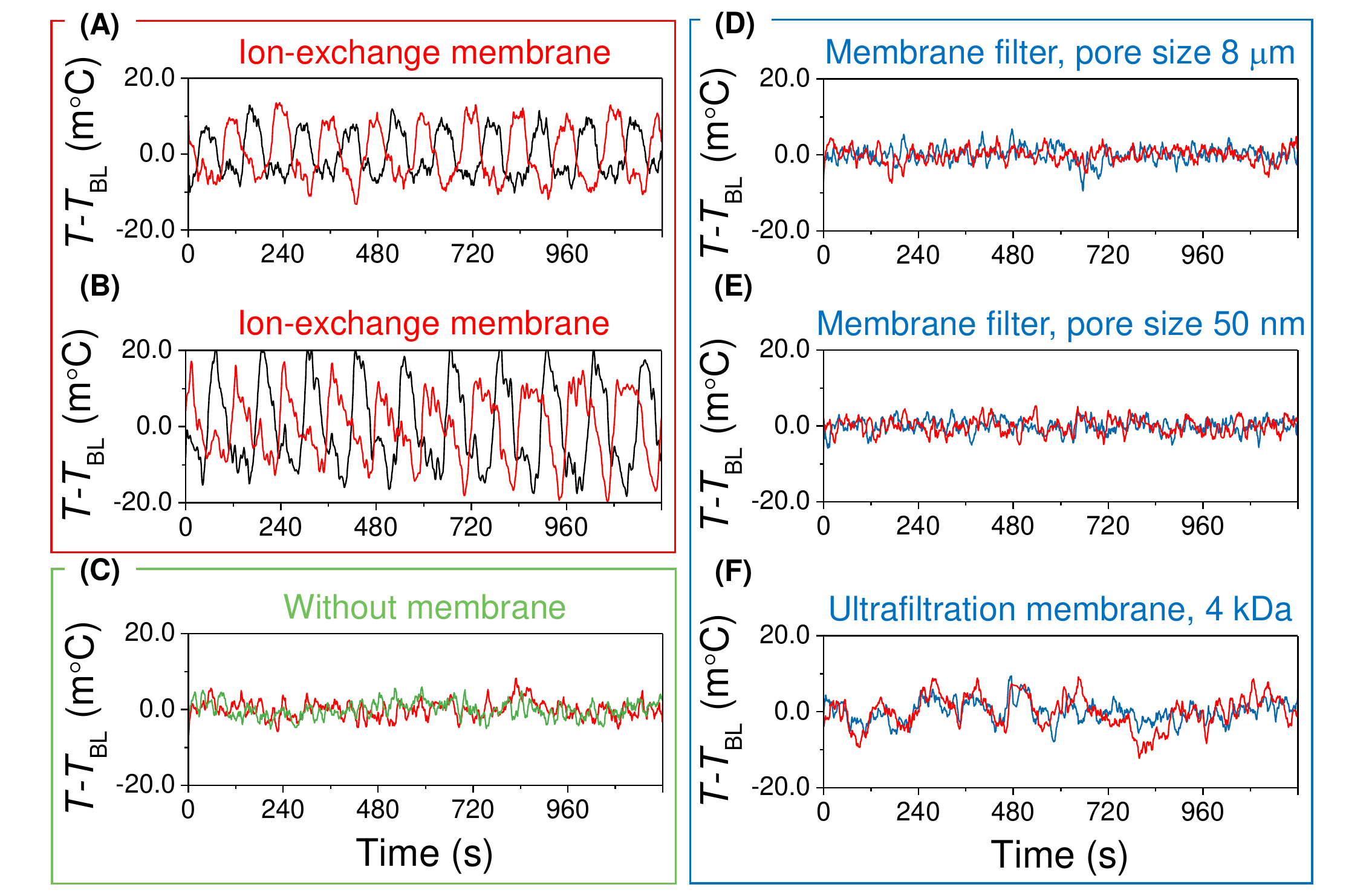}
\caption{\temphl{Current switching experiments ($I=250$ A/m$^2$, reversal of current every 60 s), with a cation exchange membrane (A,B), without a membrane (C), and with three uncharged filters or membranes (D-F).}} \label{fig5}
\end{figure}

\temphl{To support our claim that in a charged membrane (ion-exchange membrane) there is local heating and cooling in the EDLs at its surface, dependent on the membrane charge and current direction, we performed additional experiments without a membrane, and with uncharged membranes, see Fig.~\ref{fig5}, where we show the temperature traces measured by the two sensors that are placed on each side of the membrane. As before, the baseline due to the gradual temperature increase, see Fig.~\ref{fig2}A, is subtracted from the data. In Fig.~\ref{fig5}, panel A shows the same data as already presented in Fig.~\ref{fig2}B ($I=250$ A/m$^2$, switching of the current every 60 s), while panel B is a repeat experiment with a new membrane and new temperature sensors (all of the same type as before). We see in panel B the same countercyclic temperature changes on the two sides of the membrane as in panel A, though the amplitude of the signal is about 50\% higher. Using these new sensors, panel C presents the temperature profiles without any membrane, and panels D through F data for filter materials/membranes with decreasing pore sizes. Without a membrane, and for filter materials with a pore size approx. 8 $\mu$m (panel D) and approx. 50 nm (panel E), we do not observe temperature oscillations. For the membrane with the smallest pore size (a membrane that blocks molecules heavier than those with a molar mass of approx. 4 kg/mol, panel F) the temperature signal has larger variations around the baseline, but this signal is not correlated with the 60 s half-cycle time, and does not have any other clear pattern. The two sensors mostly have a synchronous time response, which is in stark contrast to the countercyclic and very repetitive behavior found in charged ion-exchange membranes (panels A and B). These additional experiments underpin our claim that in charged membranes (ion-exchange membranes) with EDLs formed on the outer surfaces, there is a source of heating and cooling that does not occur in uncharged membranes. (Details of these three filter materials/membranes are as follows: panel D: MF-Millipore Membrane Filter (Merck Millipore) consisting of a mixture of cellulose acetate and cellulose nitrate, with a porosity of 84 \% and a thickness of 135 $\mu$m; panel E: Isopore Membrane Filter (Merck Millipore) made of polycarbonate, porosity 5-20 \% and thickness 25 $\mu$m; panel F: Nadir UH004-P (Microdyn-Nadir) made of polyethersulfone and polypropylene, thickness 250 $\mu$m.)
}

In conclusion, we measured the temperatures on each side of an ion-exchange membrane through which current runs. 
We attribute the temperature difference (and its response to switching the current direction) to electrostatic cooling at one of the membrane-solution interfaces, a junction between two electrolyte phases. We can reproduce the observed switching of temperature in a theory that includes the electrical power density described by the inner product $\textbf{I}\cdot \textbf{E}$, a term which predicts cooling when the ionic current flows in a direction of increasing electrical potential (against the electric field) as occurs in one of the electrical double layers (EDLs) on the membrane surface. \temphl{When the same experiment is repeated with uncharged membranes, where EDLs are not formed on the outside surfaces, the oscillations in temperature which are observed for charged ion-exchange membranes (with the two sensors on each side of the membrane responding in a countercyclic and repetitive manner) are not found.} Our results contribute to our understanding of the structure of the EDL at the interface of membranes that are used in water treatment and desalination, as well as for energy recovery from water salinity differences. 
In addition, heating and cooling during nerve signal transduction 
also relates to ion transport and charged membranes and electrostatic heating and cooling 
can also play an important role here~\cite{Abbott_1958,Coster_1975,Ritchie_1985,Andersen_2009,Lichtervelde_2019}.

\subsection*{Acknowledgments}
This work was performed in the cooperation framework of Wetsus, European Centre of Excellence for Sustainable Water Technology (www.wetsus.eu). Wetsus is co-funded by the Dutch Ministry of Economic Affairs, the Northern Netherlands Provinces, and 
the Province of Frysl{\^a}n.


\end{document}